\documentclass[preprint,showpacs,preprintnumbers,amsmath,amssymb]{revtex4}
\usepackage{amsmath}

\usepackage{graphicx}
\usepackage{pdfpages}
\usepackage{dcolumn}
\usepackage{bm}
\usepackage{amsmath} 
\usepackage{amsfonts}
\usepackage{amssymb}
\usepackage{url}
\usepackage{microtype}
\usepackage{graphicx}%

\newcommand{\qed}{\nobreak \ifvmode \relax \else
      \ifdim\lastskip<1.5em \hskip-\lastskip
      \hskip1.5em plus0em minus0.5em \fi \nobreak
      \vrule height0.75em width0.5em depth0.25em\fi}

\renewcommand{\vec}[1]{\mathbf{#1}}

\begin{document}

\preprint{}
\title{Numerical and Exact Analyses of Bures and Hilbert-Schmidt Separability and PPT-Probabilities}
\author{Paul B. Slater}
 \email{slater@kitp.ucsb.edu}
\affiliation{%
Kavli Institute for Theoretical Physics, University of California, Santa Barbara, CA 93106-4030\\
}
\date{\today}
            
\begin{abstract}
We employ a quasirandom methodology, recently developed by Martin Roberts, to estimate the separability probabilities, with respect to the Bures (minimal monotone/statistical distinguishability) measure, of generic two-qubit and two-rebit states. This procedure, based on generalized properties of the golden ratio, 
yielded, in the course of almost seventeen billion iterations (recorded at intervals of five million),  two-qubit estimates 
 repeatedly close to nine decimal places to
$\frac{25}{341} =\frac{5^2}{11 \cdot 31} \approx 0.073313783$.  Howeer, despite the use of over twenty-three  billion
iterations,  we do not presently perceive an exact value (rational or otherwise) for an estimate of 0.15709623 for the Bures two-rebit separability probability. 
The Bures qubit-qutrit case--for which Khvedelidze and Rogojin gave an estimate of 0.0014--is analyzed too. The value of $\frac{1}{715}=\frac{1}{5 \cdot 11 \cdot 13} \approx 0.00139860$ is a well-fitting value to an estimate of 0.00139884. Interesting values ($\frac{16}{12375} =\frac{4^2}{3^2 \cdot 5^3 \cdot 11}$ and $\frac{625}{109531136}=\frac{5^4}{2^{12} \cdot 11^2 \cdot 13 \cdot 17}$) are conjectured for the Hilbert-Schmidt (HS) and Bures qubit-qudit ($2 \times 4$) positive-partial-transpose (PPT)-probabilities. We re-examine, strongly supporting, conjectures that the HS qubit-{\it qutrit} and rebit-{\it retrit} separability probabilities are 
$\frac{27}{1000}=\frac{3^3}{2^3 \cdot 5^3}$ and $\frac{860}{6561}= \frac{2^2 \cdot 5 \cdot 43}{3^8}$, respectively. Prior studies have demonstrated that the HS two-rebit separability probability is $\frac{29}{64}$ and strongly pointed to the HS two-qubit counterpart being $\frac{8}{33}$, and a certain operator monotone one (other than the Bures) being $1 -\frac{256}{27 \pi^2}$.
\end{abstract}


\maketitle
\section{Introduction}
It has now been formally proven by Lovas and Andai \cite[Thm. 2]{lovas2017invariance} that the separability probability with respect to Hilbert-Schmidt  (flat/Euclidean/Frobenius) measure \cite{zyczkowski2003hilbert} \cite[sec. 13.3]{bengtsson2017geometry} of the 9-dimensional convex set of 
two-rebit states \cite{Caves2001} is $\frac{29}{64}=\frac{29}{2^6}$. 
(``For quantum mechanics defined over real vector spaces the simplest composite systems are two-rebits systems" \cite{batle2003understanding}.) Additionally, the multifaceted evidence  \cite{slater2017master,khvedelidze2018generation,milz2014volumes,fei2016numerical,shang2015monte,slater2013concise,slater2012moment,slater2007dyson}--including a recent ``master'' extension \cite{slater2017master,slater2018extensions} of the Lovas-Andai framework to {\it generalized} two-qubit states--is strongly compelling that the corresponding value for the 15-dimensional convex set of two-qubit states is $\frac{8}{33}=\frac{2^3}{3 \cdot 11}$ (with that of the 27-dimensional convex set of two-quater[nionic]bits being $\frac{26}{323}=\frac{2 \cdot 13}{17 \cdot 19}$ [cf. \cite{adler1995quaternionic}], among other still higher-dimensional companion random-matrix related results). A still further extension to the use of induced measures--reducing to the Hilbert-Schmidt case for the case $k=0$--has been found \cite[sec. XII]{slater2018extensions}--yielding, for example, $\frac{61}{143}$ for $k=1$. 
(The parameter $k$ is the difference [$k= K-N$] between the dimensions [$K,N$,with $K\geq N$] of the subsystems of the pure state bipartite system in which the density matrix is regarded as being embedded \cite{zyczkowski2001induced}.)

Further, appealing hypotheses parallel to these rational-valued results have been advanced--based on extensive sampling--that the Hilbert-Schmidt separability probabilities for the 35-dimensional qubit-{\it qutrit} and 20-dimensional rebit-{\it retrit} states are 
$\frac{27}{1000}=\frac{3^3}{2^3 \cdot 5^3}$ and $\frac{860}{6561}= \frac{2^2 \cdot 5 \cdot 43}{3^8}$, respectively \cite[eqs. (15),(20)]{slater2018extensions} \cite[eq. (33)]{milz2014volumes}. (These will be further examined in sec.~\ref{HSsection} below.)

Certainly, one can, however, still aspire to a yet greater ``intuitive'' understanding of these assertions, particularly in some ``geometric/visual'' sense 
[cf. \cite{szarek2006structure,samuel2018lorentzian,avron2009entanglement,braga2010geometrical,gamel2016entangled,jevtic2014quantum}], as well as further formalized proofs. It would be of interest, as well,  to compare/contrast these finite-dimensional studies with those other quantum-information-theoretic ones, presented in the recent comprehensive volume of Aubrun and Szarek
\cite{aubrun2017alice}, in which the quite different concepts of {\it asymptotic geometric analysis} are employed.

By a separability probability, in the above discussion, we mean the ratio of the volume of the separable states to the volume  of all (separable and entangled) states, as proposed, apparently first, by {\.Z}yczkowski, Horodecki, Sanpera and Lewenstein \cite{zyczkowski1998volume} (cf. \cite{petz1996geometries,e20020146,singh2014relative,batle2014geometric}). The present author was, then, led--pursuing an interest in ``Bayesian quantum mechanics" \cite{slater1994bayesian,slater1995quantum} and the concept of a ``quantum Jeffreys prior" \cite{kwek1999quantum}--to investigate how such separability probabilities might depend upon the choice of various possible measures on the quantum states \cite{petz1996geometries}.
\subsection{Partitioning of separability/PPT-probabilities}
\subsubsection{Hilbert-Schmidt and Bures cases}
A phenomenon apparently restricted to the Hilbert-Schmidt ($k=0$) case of induced measure, is that the positive-partial-transpose (PPT) states are equally divided  probability-wise between those for which the determinant $|\rho^{PT}|$ of the partial transpose of the density matrix ($\rho$) exceeds the determinant $|\rho|$ of the density matrix itself and {\it vice versa}. (Also, along somewhat similar lines, the Hilbert-Schmidt PPT-probability for minimally degenerate [having a single zero eigenvalue] states is {\it half} that for nondegenerate states \cite{szarek2006structure}. The PPT-property is, of course, equivalent--by the Peres-Horodecki criterion--to separability for $4  \times 4$ and $6 \times 6$ density matrices \cite[sec. 16.6.C]{bengtsson2017geometry}.) Quite contrastingly, based on some 122,000,000 two-qubit density matrices randomly generated with respect 
to Bures measure, of the 8,945,951 separable ones found, 5,894,648 of them (that is, 65.89$\%$) had $|\rho^{PT}| >|\rho|$, clearly distinct from simply 50$\%$ (cf. \cite[Tabs. 1, 2]{khvedelidze2018generation} \cite{KhvRog}).
\subsubsection{Induced measures, in general}
A formula   
for that part, $Q(k,\alpha)$, of the {\it total} separability probability, $P_{sep}(k,\alpha)$,
for generalized (real [$\alpha=1$], complex [$\alpha=2$], quaternionic [$\alpha=4$],\ldots) two-qubit states endowed with random induced measure, for which   the determinantal inequality $|\rho^{PT}| >|\rho|$ holds was given in \cite[p. 26]{slater2016formulas}. It took the form $Q(k,\alpha)= G_1^k(\alpha) G_2^k(\alpha)$, for $k = -1, 0, 1,\ldots 9$. 
(The factors $G^k_2(\alpha)$ are sums of polynomial-weighted generalized hypergeometric functions $_pF_{p-1}, p \geq 7$, all with argument $z = \frac{27}{64}$.)
Here $\rho$ denotes a $4 \times 4$ density matrix, obtained by tracing over the pure states in $4 \times (4 +k)$-dimensions, and $\rho^{PT}$, its partial transpose. Further, $\alpha$ is a Dyson-index-like parameter with $\alpha = 1$ for the standard (15-dimensional) convex set of (complex) two-qubit states.   

Further, in the indicated ($k=0$) Hilbert-Schmidt case, we can apparently employ 
the formula \cite[p. 26]{slater2016formulas}
\begin{equation} \label{InducedMeasureCase}
\mathcal{P}_{sep/PPT}(0,\alpha)= 2 Q(0,\alpha)= 1-    
\frac{\sqrt{\pi } 2^{-\frac{9 \alpha}{2}-\frac{5}{2}} \Gamma \left(\frac{3 (\alpha+1)}{2}\right)
   \Gamma \left(\frac{5 \alpha}{4}+\frac{19}{8}\right) \Gamma (2 \alpha+2) \Gamma \left(\frac{5
   \alpha}{2}+2\right)}{\Gamma (\alpha)} \times
\end{equation}
\begin{displaymath}
\, _6\tilde{F}_5\left(1,\alpha+\frac{3}{2},\frac{5 \alpha}{4}+1,\frac{1}{4} (5 \alpha+6),\frac{5
   \alpha}{4}+\frac{19}{8},\frac{3 (\alpha+1)}{2};\frac{\alpha+4}{2},\frac{5
   \alpha}{4}+\frac{11}{8},\frac{1}{4} (5 \alpha+7),\frac{1}{4} (5 \alpha+9),2 (\alpha+1);1\right).
\end{displaymath}
That is, for $k=0$, we obtain the previously reported Hilbert-Schmidt formulas, with
(the real case) $Q(0,1) = \frac{29}{128}$, (the standard complex case) $Q(0,2)=\frac{4}{33}$, and 
(the quaternionic case) $Q(0,4)= \frac{13}{323}$---the three  simply 
equalling--by the equipartitioning result noted above--$ P_{sep}(0,\alpha)/2$.
More generally, $Q(k,\alpha)$ gives that portion, for  induced measure, parameterized by $k$, of the total separability/PPT-probability for which the determinantal inequality 
$|\rho^{PT}| >|\rho|$ holds \cite[eq. (84)]{slater2017master}. 
\section{Use of Bures Measure}
Of particular initial interest was the the Bures/statistical distinguishability (minimal monotone) measure \cite{slater2000exact,sarkar2019bures,vsafranek2017discontinuities,forrester2016relating, braunstein1994statistical}. (``The Bures metric 
plays a distinguished role since it is the only metric which is also monotone, 
Fisher-adjusted, Fubini-Study-adjusted, and Riemannian" \cite{forrester2016relating}. Bej and Deb have recently ``shown that if a qubit gets entangled with another ancillary qubit then negativity, up to a constant factor, is equal to square root of a specific Riemannian metric defined on the metric space corresponding to the state space of the qubit" \cite{bej2018geometry}.)

In \cite[sec. VII.C]{slater2017master}, we recently reported, building upon analyses of Lovas and Andai \cite[sec. 4]{lovas2017invariance}, a two-qubit separability probability equal to $1 -\frac{256}{27 \pi^2} =1- \frac{2^8}{3^3 \pi^2} \approx 0.0393251$. This was based on another 
(of the infinite family of) operator monotone functions, namely 
$\sqrt{x}$. (The Bures measure itself is associated with the operator monotone function $\frac{1+x}{2}$.) (Let us note that the complementary ``entanglement probability'' is simply $\frac{256}{27 \pi^2} \approx 0.960675$. There appears to be no intrinsic reason
to prefer/privilege one of these two forms (separability, entanglement) of probability to the other [cf. \cite{dunkl2015separability}].  We observe that the  variable  denoted $K_s = \frac{(s+1)^{s+1}}{s^s}$, equalling $\frac{256}{27} = \frac{4^4}{3^3}$, for $s=3$, is frequently employed as an upper limit of integration in the Penson-{\.Z}yczkowski paper, ``Product of Ginibre matrices: Fuss-Catalan and Raney 
distributions'' \cite[eqs. (2),  (3)]{penson2011product}.)  

Interestingly, Lovas and Andai ``argue that from the separability probability point of view, the main difference between the Hilbert-Schmidt measure and the volume form
generated by the operator monotone function $x \rightarrow \sqrt{x}$ is a special distribution on the unit ball in operator norm of 
$2 \times 2$ matrices, more precisely in the Hilbert-Schmidt case one faces a uniform distribution on the whole unit ball and for 
monotone volume forms one obtains uniform distribution on the surface of the unit ball'' \cite[p. 2]{lovas2017invariance}. 
\subsection{Osipov-Sommers-{\.Z}yzckowski Interpolation Formula}
Of central importance in our analyses below will be  the construction of Osipov, Sommers and {\.Z}yzckowski of an interpolation between the generation of random 
density matrices with respect to Hilbert-Schmidt and those with respect to Bures measures \cite[eq. (24)]{al2010random} (cf. \cite[eq. (33)]{borot2012purity}). This formula takes the form
\begin{equation}  \label{JointBuresHSformula}
\rho_x= \frac{(y \mathbb{I} +x U) AA^{\dagger}(y \mathbb{I}+x U^{\dagger})}{\mbox{Tr} (y \mathbb{I}+x U) A A^{\dagger} (y \mathbb{I}   +x U^{\dagger})},
\end{equation}
where $y=1-x$, with $x=0$ yielding a Hilbert-Schmidt density matrix $\rho_0$, and $x=\frac{1}{2}$, the Bures counterpart $\rho_{1/2}$. 
Here, $A$ is an $N \times N$ complex-valued random matrix pertaining to the Ginibre ensemble (with real and imaginary parts of each of the $N^2$ entries being independent standard normal random variates). Further, $U$ is a random
unitary matrix  distributed according to the Haar measure on $U(N)$. (Of course, $N=4$ in the basic two-qubit case of first interest here.)

It is an intriguing hypothesis that the Bures two-qubit separability probability  also assumes a strikingly elegant form (such as the indicated $\frac{8}{33}$, $1-\frac{256}{27 \pi^2}$). (``Observe that the Bures volume of the set of mixed states is equal to the volume of an $(N^2-1)$-dimensional hemisphere of radius $R_B=\frac{1}{2}$'' \cite[p. 415]{bengtsson2017geometry}. It is also noted there that $R_B$ times the area-volume ratio asymptotically increases with the dimensionality $D=N^2-1$, which is typical for hemispheres.)
\subsection{Prior Estimations of Bures separability probabilities}
In the relatively early (2002) work 
\cite{slater2002priori}, we had conjectured a Bures two-qubit separability probability equal to $\frac{8}{11 \pi^2} \approx .0736881$. But it was later proposed   
in 2005 \cite{slater2005silver}, in part motivated by the lower-dimensional {\it exact} Bures results reported in \cite{slater2000exact}, that the value might be $\frac{1680 \sigma_{Ag}}{\pi^8} \approx 0.07334$, where $\sigma_{Ag}= \sqrt{2}-1 \approx 0.414214$ is the ``silver mean''. Both of these studies \cite{slater2002priori,slater2005silver}   were conducted using quasi-Monte Carlo procedures, before the developmentof the indicated Osipov-Sommers-{\.Z}yczkowski methodology (\ref{JointBuresHSformula}) for generating density matrices, random with respect to  Bures measure \cite{al2010random}. More recently, in \cite[sec. X.B.1]{slater2018extensions}, we reported, using this Ginibre ensemble-based formula (\ref{JointBuresHSformula}) an estimate of 0.0733181043 based on 4,372,000,000 realizations, using simply standard
(independent) random normal variate generation. (Khvedelidze and Rogojin 
gave a value of 0.0733 \cite[Table 1]{khvedelidze2018generation} \cite{KhvRog}.)

Performing a parallel (but much smaller) computation in the two-rebit case, based on forty million random density matrices (6,286,209 of them being separable), we obtained a corresponding (slightly corrected) Bures separability probability estimate of 0.1571469. (In doing so, we took, as required, the now real-entried Ginibre matrix A  to be $4 \times 5$ \cite[eqs. (24), (28)]{al2010random}, and not $4 \times 4$ as in the two-qubit calculation.)

\subsection{Application of Quasirandom Methodology to Bures Two-Rebit and Two-Qubit Cases}
We now importantly examine the question of whether   Bures two-qubit and two-rebit separability probability estimation can be accelerated--with superior convergence properties--by, rather than using, as typically done, 
{\it independently}-generated normal variates for the Ginibre ensembles at each iteration, making use of normal variates {\it jointly}-generated by employing low-discrepancy (quasi-Monte Carlo) sequences \cite{leobacher2014introduction}. In particular, we have employed an ``open-ended'' sequence (based on extensions of the  golden ratio \cite{livio2008golden}) recently introduced by Martin Roberts in the detailed presentation ``The Unreasonable Effectiveness 
of Quasirandom Sequences'' \cite{Roberts}.
 
Roberts notes: ``The solution to the 
$d$-dimensional problem, depends on a special constant $\phi_d$, where $\phi_d$ is the value of the smallest, positive real-value of x such that''
\begin{equation}
  x^{d+1}=x+1,
\end{equation}
($d=1$, yielding the golden ratio, and $d=2$, the ``plastic constant'' \cite{Roberts32D}). 
The  $n$-th terms in the quasirandom (Korobov) sequence take the form
\begin{equation} \label{QR}
(\alpha _0+n \vec{\alpha}) \bmod 1, n = 1, 2, 3, \ldots  
\end{equation}
where we have the $d$-dimensional vector,
\begin{equation} \label{quasirandompoints}
\vec{\alpha} =(\frac{1}{\phi_d},\frac{1}{\phi_d^2},\frac{1}{\phi_d^3},\ldots,\frac{1}{\phi_d^d}).  "
\end{equation}
The additive constant $\alpha_0$ is typically taken to be 0. ``However, there are some arguments, relating to symmetry, that suggest that $\alpha_0=\frac{1}{2}$
is a better choice,''  Roberts observes. 

These points (\ref{QR}), {\it uniformly} distributed in the $d$-dimensional hypercube $[0,1]^d$, can be converted to (quasirandomly distributed) normal variates, required for implementation of the Osipov-Sommers-{\.Z}yczkowski formula (\ref{JointBuresHSformula}), using the inverse of the cumulative distribution function \cite[Chap. 2]{devroye1986}. 
Impressively, in this regard, Henrik Schumacher developed for us a specialized algorithm that accelerated the  default Mathematica command InverseCDF for the normal distribution approximately {\it ten-fold}, as reported in the highly-discussed post \cite{Schumacher}--allowing us to vastly increase the realization rate.

We take $d=36$ and 64 in the Roberts methodology, using the Osipov-Sommers-{\.Z}yczkowski (real and complex) interpolation formulas to estimate the Bures two-rebit and two-qubit separability probabilities, respectively. In the two-qubit case, 32 of the 64 variates are used in generating the Ginibre matrix $A$, and the other 32, for the unitary matrix $U$. 
(A subsidiary question--which appeared in the discussion with Roberts \cite{Roberts32D}--is the relative effectiveness of employing--to avoid possible ``correlation'' effects--the {\it same} 32-dimensional sequence but at different $n$'s for $A$ and $U$, rather than a single 64-dimensional one, as pursued here. A small analysis of ours in this regard did not indicate this to be a meritorious approach.) In the two-rebit case, 20 variates are used to generate 
the $4 \times 5$ matrix A, and the other 16 for an orthogonal $4 \times 4$ matrix $O$.

\begin{figure}
\includegraphics[]{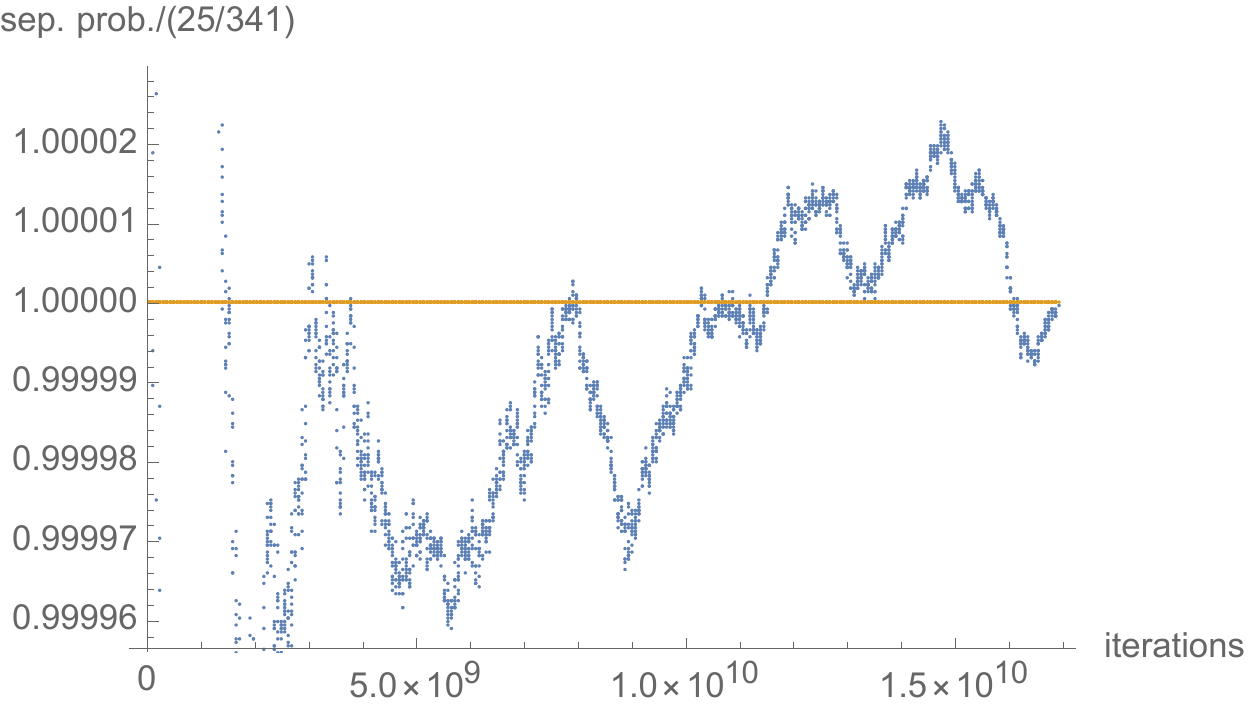}
\caption{Two-qubit Bures separability probability estimates--divided by $\frac{25}{341}$--as a function of the number of iterations of the quasirandom procedure, using $\alpha_0=\frac{1}{2}$. This ratio is equal to 1 to nearly eight decimal places at: 1,445,000,000; 10,850,000,000; 11,500,000,000; and 16,075,000,000 iterations. Estimates are recorded at intervals of five million iterations.}
\label{fig:twoqubitplot}
\end{figure}

\begin{figure}
\includegraphics[]{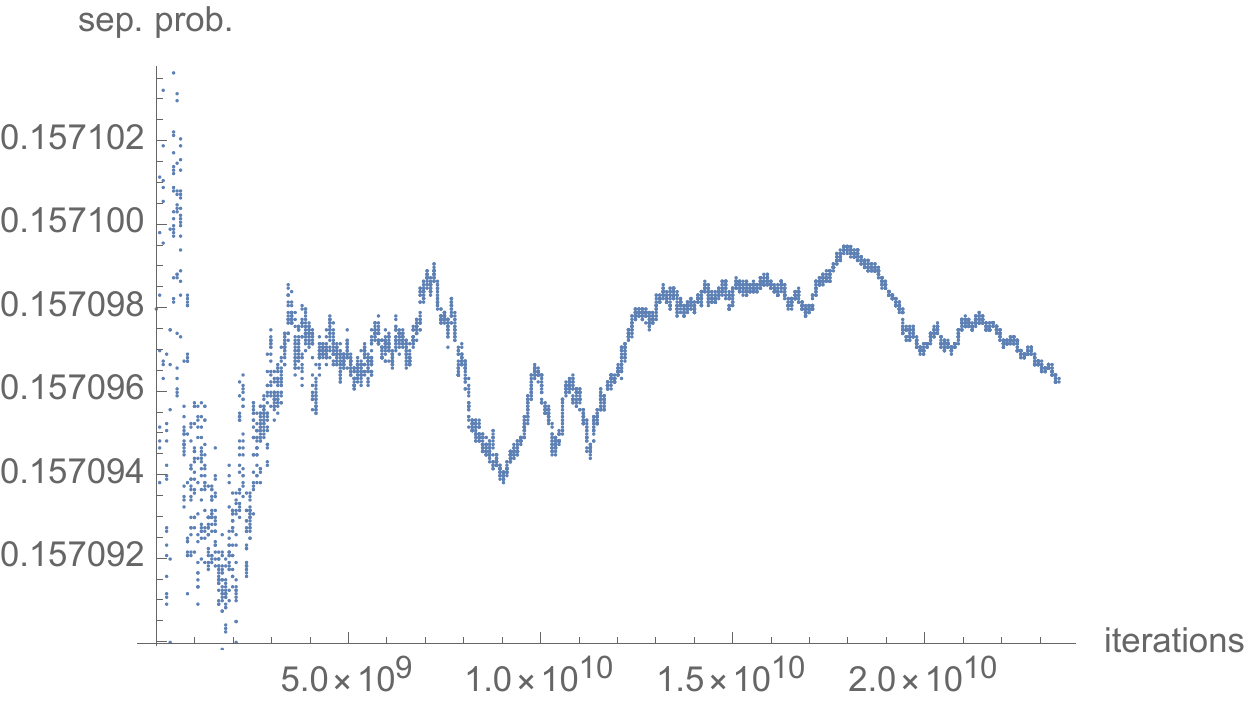}
\caption{Two-rebit Bures separability probability estimates as a function of the number of iterations of the quasirandom procedure, using $\alpha_0=\frac{1}{2}$. Estimates are recorded at intervals of five million iterations.}
\label{fig:tworebitplot}
\end{figure}

In Figs.~\ref{fig:twoqubitplot} and \ref{fig:tworebitplot} we show the development of the Bures separability probability estimation procedure in the two cases at hand. (Much earlier versions of these [$\alpha_0=\frac{1}{2}$] plots--together with [less intensive] estimates using $\alpha_0=0$--were displayed as Figs. 5 and 6 in \cite{slater2018extensions}.)
\subsubsection{Two-qubit Bures analysis}
Using the indicated, possibly superior parameter value 
$\alpha_0= \frac{1}{2}$  in (\ref{QR}), this quasirandom/normal-variate-generation procedure 
yielded a two-qubit estimate, based on 16,895,000,000 iterations, of 0.073313759. This is  closely fitted by the two (themselves very near) values
$\frac{25}{341} =\frac{5^2}{11 \cdot 31} \approx 0.07331378299$ and (as suggested by the WolframAlpha.com site) $\frac{\sqrt{51}}{\pi ^4} \approx 0.07331377752$. (Informally, Charles Dunkl wrote: "I would hate to think that the answer is $\frac{\sqrt{51}}{\pi^4}$- that is just ugly. One hopes for a rational number.") At 10,850,000,000 iterations, interestingly, the estimate of 0.0733137814 matched $\frac{25}{341}$ to nearly eight decimal places. The estimate of  0.0733137847 obtained at the considerably smaller number of iterations of 1,445,000,000, was essentially as close too. 
The Hilbert-Schmidt measure counterpart is (still subject to formal proof) essentially known to be $\frac{8}{33} = \frac{2^3}{3 \cdot 11}$ \cite{slater2017master,khvedelidze2018generation,milz2014volumes,fei2016numerical,shang2015monte,slater2013concise,slater2012moment,slater2007dyson}. 
\subsubsection{Two-rebit Bures analysis} \label{BuresTwoRebit}
In the two-rebit instance, we obtained a Bures estimate, based on 23,460,000,000 
iterations, of 0.157096234. This is presumably, at least as accurate as the considerably, just noted, smaller sample based two-qubit estimate--apparently corresponding to $\frac{25}{341}$. Nevertheless, we do not presently perceive any possible exact--rational or otherwise--fits to this estimate. 

While, the Hilbert-Schmidt two-rebit separability probability 
has been proven by Lovas and Andai to be $\frac{29}{64} = \frac{29}{2^6}$ \cite[Thm. 2]{lovas2017invariance}, somewhat similarly to this Bures result, the two-rebit separability probability, 0.2622301318,  based on the other monotone ($\sqrt{x}$) measure, did not seem to have an obvious exact underlying formula.  
\section{Examination of Hilbert-Schmidt Qubit-Qutrit and Rebit-Retrit Separability Conjectures} \label{HSsection}
\subsection{Prior studies}
Based on extensive (standard) random sampling of independent normal variates, in \cite[eqs. (15),(20)]{slater2018extensions}, we  have conjectured that the Hilbert-Schmidt separability probabilities for the 35-dimensional qubit-{\it qutrit} and 20-dimensional rebit-{\it retrit} states are  (also interestingly rational-valued) 
$\frac{27}{1000}=\frac{3^3}{2^3 \cdot 5^3} =0.027$ and $\frac{860}{6561}= \frac{2^2 \cdot 5 \cdot 43}{3^8} \approx 0.1310775796$, respectively . 
In particular, on the basis of 2,900,000,000 randomly-generated 
qubit-qutrit density matrices, an estimate (with 78,293,301
separable density matrices found) was obtained, yielding an associated separability probability of 0.026997690. (Milz and Strunz had given a confidence interval of $0.02700 \pm 0.00016$ for this probability \cite[eq. (33)]{milz2014volumes}, while Khvedelidze and Rogojin reported an estimate of 0.0270 \cite[Tab. 1]{khvedelidze2018generation}--but also only 0.0014 for the Bures counterpart [sec.~\ref{BuresQubitQutrit}].)
Further, on the basis of 3,530,000,000 randomly-generated 
rebit-retrit density matrices, with respect to Hilbert-Schmidt measure, an estimate (with 462,704,503
separable density matrices found) was obtained for an associated separability probability of 0.1310777629. The associated 
$95\%$ confidence interval is $[0.131067, 0.131089]$.
\subsection{New studies}
\begin{figure}
    \centering
    \includegraphics{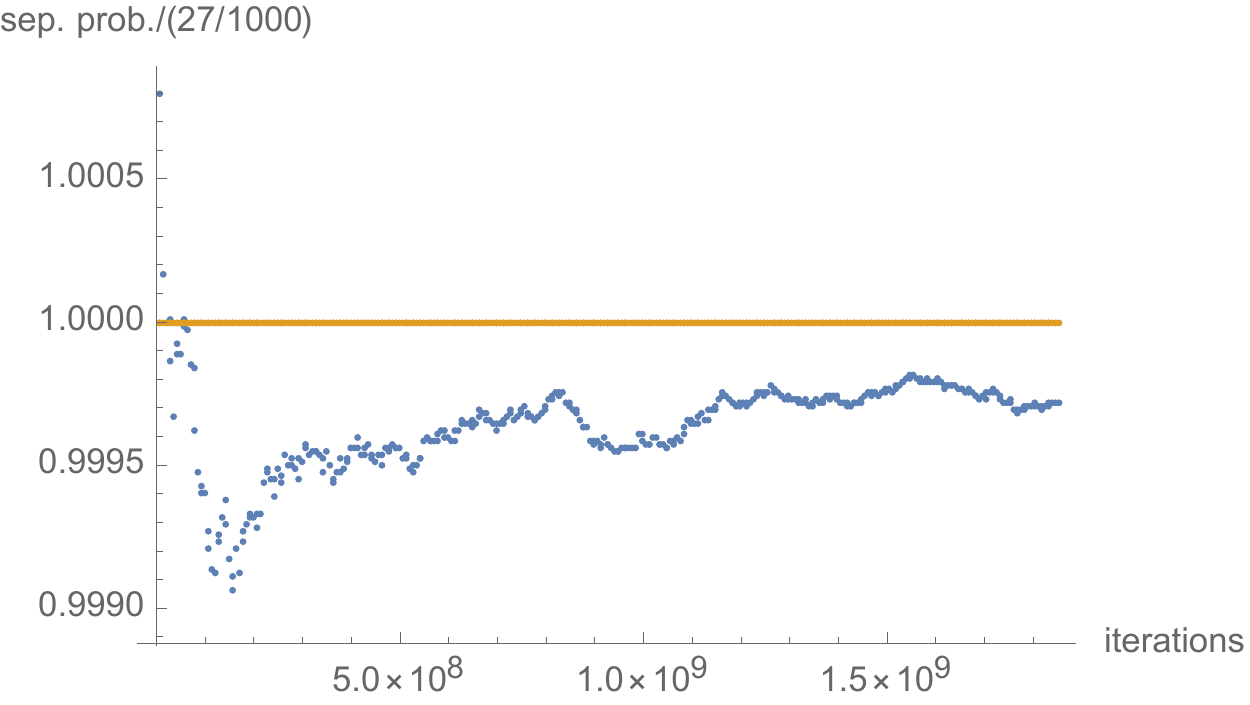}
    \caption{Qubit-qutrit Hilbert-Schmidt separability probability estimates--divided by $\frac{27}{1000}$--as a function of the number of iterations of the quasirandom procedure, using $\alpha_0=\frac{1}{2}$. Estimates are recorded at intervals of five million iterations.}
    \label{fig:QuasiRandomQubitQutrit}
\end{figure}

\begin{figure}
    \centering
    \includegraphics{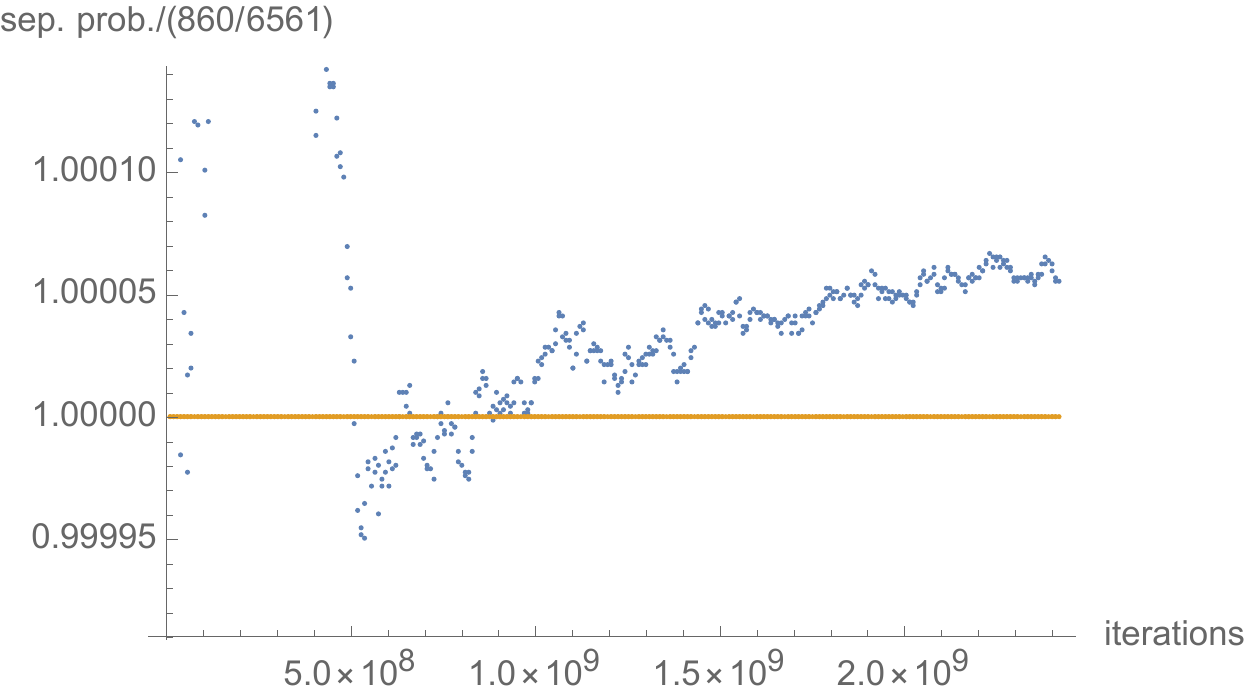}
    \caption{Rebit-retrit Hilbert-Schmidt separability probability estimates--divided by $\frac{860}{6561} = \frac{2^2 \cdot 5 \cdot 43}{3^8}$--as a function of the number of iterations of the quasirandom procedure, using $\alpha_0=\frac{1}{2}$. Estimates are recorded at intervals of five million iterations.}
    \label{fig:QuasiRandomRebitRetrit}
\end{figure}

Applying the quasirandom methodology here to further appraise this pair of conjectures, we obtain Figs.~\ref{fig:QuasiRandomQubitQutrit} and \ref{fig:QuasiRandomRebitRetrit}.  (We take the dimensions $d$ of the sequences of normal variates generated to be 72 and 42, respectively.)
Interestingly, as in Fig.~\ref{fig:twoqubitplot}, we observe some 
drift away--with increasing iterations--from early particularly close fits
to the two conjectures. But, as in Fig.~\ref{fig:twoqubitplot}--assuming the validity
of the conjectures--we might anticipate the estimates re-approaching more closely
the conjectured values. It would seem that any presumed eventual convergence is not simply a
straightforward monotonic process--perhaps somewhat comprehensible in view of the very high dimensionalities (72, 42) of the sequences involved. (The last recorded separability probabilities--in these ongoing analyses--were 0.0269923 and 0.1310848, based on 1,850,000,000 and 
2,415,000,000 iterations, respectively.)

In \cite[App. B]{slater2017master}, we reported an effort to extend the innovative framework of Lovas and Andai \cite{lovas2017invariance} to such qubit-qutrit and rebit-retrit  settings. (One aspect of interest pertaining to the original $4 \times 4$ density matrix study of Lovas and Andai \cite{lovas2017invariance}, was that it (surprisingly) appeared possible in \cite{slater2017master} to extend the original Lovas-Andai framework by restricting our analyses to $4 \times 4$ density matrices in which the two $2 \times 2$ diagonal blocks were themselves diagonal.)

\section{An 8-dimensional ($X$-states) rebit-retrit scenario} \label{8D}
Along similar lines, let us consider an 8-dimensional ($X$-states) rebit-retrit scenario, in which now the only non-zero entries of $\rho$ are those on the diagonal and anti-diagonal--so that the two $3 \times 3$ diagonal blocks are themselves diagonal. Also, let us employ the "separability function" framework developed in 
\cite[eq. (5)]{slater2008extended}, where the variable 
$\eta= \frac{\rho_{11} \rho_{66}}{\rho_{33} \rho_{44}}$
was employed.

Then, with the use of the Mathematica command GenericCylindricalDecomposition--employed to enforce the positivity of leading minors of the density matrix and its partial transpose--we are able to formally establish that the associated rebit-retrit Hilbert-Schmidt separability probability is $\frac{16}{3 \pi^2} \approx 0.54038$  \cite{dunkl2015separability}. 
(This value also holds for the two-rebit and two-retrit $X$-states, while $\frac{2}{5}$ is the two-qubit $X$-states probability  
\cite[sec. VIII]{slater2018extensions}.) The value $\frac{16}{3 \pi^2}$ is obtained--through integration using the output of this Mathematica command--by taking the ratio of 
\begin{equation} \label{8dNumerator}
\int_{\eta=0}^1 \frac{\pi  \eta  \left(-3 \eta ^2+(\eta +4) \eta  \log (\eta )+\log (\eta
   )+3\right)}{40320 (\eta -1)^5} d \eta =\frac{\pi}{967680}= \frac{\pi}{2^{10} \cdot 3^3 \cdot 5 \cdot 7} 
\end{equation}
to 
\begin{equation} \label{8dDenominator}
\int_{\eta=0}^1 \frac{\pi  \sqrt{\eta } \left(-3 \eta ^2+(\eta +4) \eta  \log (\eta )+\log (\eta
   )+3\right)}{40320 (\eta -1)^5} d \eta =\frac{\pi^3}{5160960} =\frac{\pi^3}{2^{14} \cdot 3^2 \cdot 5 \cdot 7},
\end{equation}
where $\sqrt{\eta}$ plays the role of separability function, and is the added factor--that is, $\eta= (\sqrt{\eta})^2$--in the first of the two integrands immediately above.
\section{Application in higher dimensions of master Lovas-Andai generalized two-qubit formulas}
We investigated extending this 8-dimensional rebit-retrit analysis to a 10-dimensional one, by replacing two previously zero entries, so that the two off-diagonal $3 \times 3$ blocks now themselves form X-patterns. The counterpart of the denominator function (\ref{8dDenominator}) is, then,
\begin{equation}
\int_{\eta=0}^1 \frac{\pi  \eta  (3 (\eta +1) (\eta  (\eta +8)+1) \log (\eta )-(\eta -1) (\eta  (11 \eta
   +38)+11))}{1209600 (\eta -1)^7}d \eta  = \frac{\pi}{29030400} = 
\end{equation}
\begin{displaymath}
   \frac{\pi}{2^{11} \cdot 3^4 \cdot 5^2 \cdot 7}.
\end{displaymath}
We, then, need to find the appropriate separability function--corresponding to $\sqrt{\eta}$ in (\ref{8dNumerator})--to insert into this integrand--for the numerator--to complete the calculation of the separability probability ratio. In this regard, we were able to, preliminarily, utilize a sub-optimal separability function (based on the enforcement of the positivity of the determinant of the $5 \times 5$ leading submatrix of the partial transpose--but not yet the full determinant),
\begin{equation}
\frac{2 \left(\sqrt{(1-\eta) \eta}+\sin ^{-1}\left(\sqrt{\eta}\right)\right)}{\pi },
\end{equation}
which yields an upper bound on the separability probability of 
$\frac{919}{5} -264 \log (2) \approx 0.809144$. 

Then--using the full determinant--we were able to construct the actual separability 
function \cite{Heidecki,Student},
\begin{equation} \label{sepfunct}
\frac{2 \left(\varepsilon ^2 \left(4 \text{Li}_2(\varepsilon )-\text{Li}_2\left(\varepsilon
   ^2\right)\right)+\varepsilon ^4 \left(-\tanh ^{-1}(\varepsilon )\right)+\varepsilon ^3-\varepsilon
   +\tanh ^{-1}(\varepsilon )\right)}{\pi ^2 \varepsilon ^2},    
\end{equation}
where the dilogarithm is indicated, and $\epsilon^2=\eta$.
The corresponding separability probability was, then, shown to be \cite{Student}
\begin{equation}
\frac{272}{45 \pi^2} \approx  0.612430.  
\end{equation}
(We have also found very strongly convincing numerical evidence that the same separability probability holds, if instead of considering ten-dimensional rebit-retrit scenarios in which the two off-diagonal $3 \times 3$ blocks have $X$-patterns, one considers that the two diagonal $3 \times 3$ blocks do.)

Further, it appears remarkable that the 10-dimensional rebit-retrit separability function (\ref{sepfunct}) turned out to be completely identical with the (polylogarithmic) Lovas-Andai two-rebit function $\tilde{\chi}_1(\varepsilon)$ \cite[eq.(2)]{slater2017master} \cite[eq. (9)]{lovas2017invariance}.

Then, in light of this finding, it appears reasonable to entertain  an hypothesis that the Lovas-Andai two-qubit function $\tilde{\chi}_2(\varepsilon)=\frac{1}{3} \varepsilon^2 (4 -\varepsilon^2)$ and two-quaterbit function $\tilde{\chi}_{4}(\varepsilon)=\frac{1}{35} \varepsilon ^4 \left(15 \varepsilon ^4-64 \varepsilon ^2+84\right)$ play parallel roles when the associated sets of density matrices share the same zero-nonzero pattern as the two ten-dimensional sets of rebit-retrit density matrices just considered (those with either the two off-diagonal or the two diagonal $3 \times 3$ blocks having $X$-patterns).

Pursuing such an hypothesis, and employing polar and ``hyper-polar'' coordinates in the very same manner as was done in \cite{slater2017master}, we can readily perform computations, in these higher-dimensional settings, leading to a presumptive qubit-qutrit separability probability of $\frac{5}{3} \left(112 \pi ^2-1105\right) = \frac{560 \pi ^2}{3}-\frac{5525}{3} \approx 0.659488$ and a quaterbit-quatertrit separability probablility of $\frac{8962661573}{4725}-192192 \pi ^2 \approx 0.583115$. (Let us interestingly note that $1105 = 5 \cdot 13 \cdot 17$, $112=2^4 \cdot 7$,$192192-2^6 \cdot 3 \cdot 7 \cdot 11 \cdot 13$ and $4725 = 3^3 \cdot 5^2 \cdot 7$. Also, we will note that $8962661573 = 193 \cdot 46438661$.)

We have tried directly computing/approximating--an apparently rather challenging task--this hypothesized  qubit-qutrit separability probability value of $\frac{5}{3} \left(112 \pi ^2-1105\right) \approx 0.659488$--that is, without simply assuming the applicability of $\tilde{\chi}_2(\varepsilon)$, but have only obtained a value of 0.67696 \cite{qubitqutrit}.

\section{Enlarged Two-Retrit $X$-states}
It has been established--as previously noted (sec.~\ref{8D})--that the Hilbert-Schmidt separability-PPT probabilities are all equal to $\frac{16}{3 \pi^2}$ for the two-rebit, rebit-retrit and two-retrit $X$-states.
Then, continuing along the lines we have just been investigating, we  considered a scenario in which the two-retrit $X$-states gained a non-zero (1,2)-entry.
Then, we, in fact,  were able to determine that the Hilbert-Schmidt PPT-probability for this scenario was $\frac{65}{36 \pi}$, making use of a separability function $\frac{8 \left(\sqrt{1-u^2} u^2-\sqrt{1-u^2}+1\right)}{3 \pi  u}$, where $u=\sqrt{\frac{\rho_{33} \rho_{77}}{\rho_{11} \rho_{99}}}$. (We found identical results when the entry chosen to be non-zero was the (1,4)--and not the (1,2)--one.)
\section{Bures Qubit-Qutrit Analysis} \label{BuresQubitQutrit}
\begin{figure}
    \centering
    \includegraphics{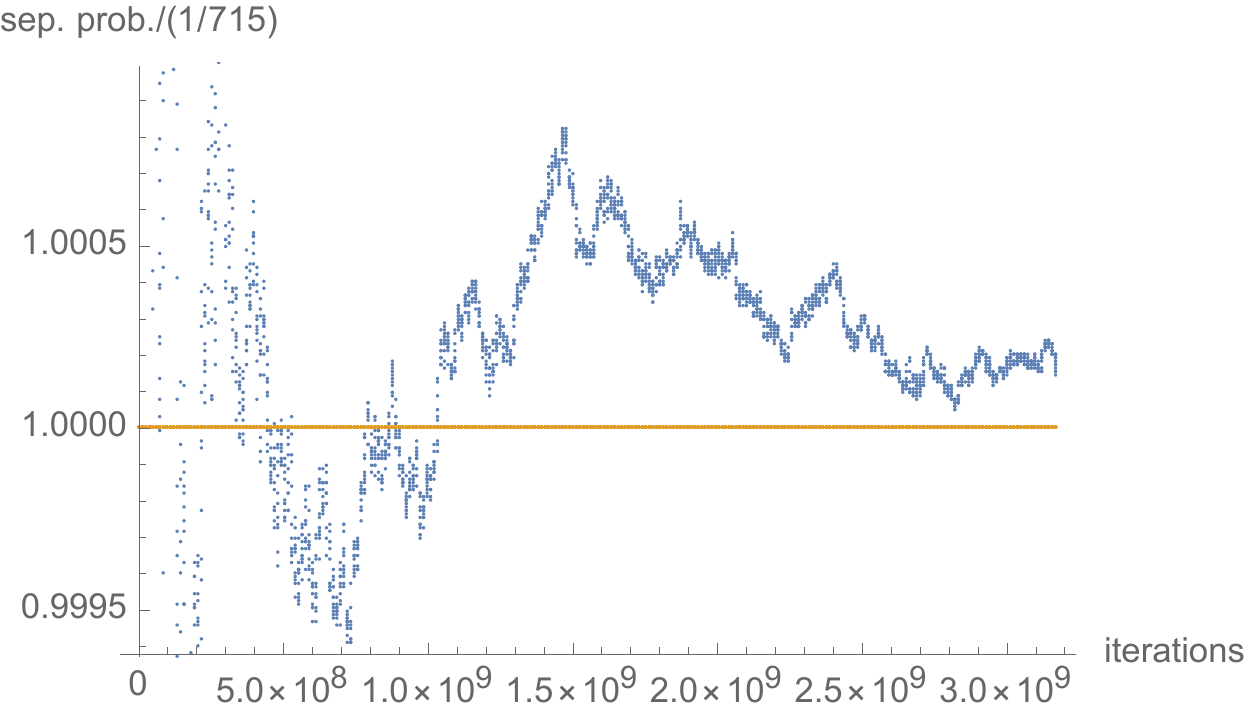}
    \caption{Qubit-qutrit Bures separability probability estimates--divided by $\frac{1}{715} = \frac{1}{5 \cdot 11 \cdot 13}$--as a function of the number of iterations of the quasirandom procedure, using $\alpha_0=\frac{1}{2}$. Estimates are recorded at intervals of one million iterations.}
    \label{fig:BuresQubitQutrit}
\end{figure}

In Table 1 of their recent study, "On the generation of random ensembles of qubits and qutrits: Computing separability probabilities for fixed rank states'' \cite{khvedelidze2018generation}, Khvedelidze and Rogojin report an estimate (no sample size being given) of 0.0014 for the separability probability of the 35-dimensional convex set of qubit-qutrit states. We undertook a study of this issue, once again employing the quasirandom methodology advanced by Roberts (with the sequence dimension parameter $d$ now equal to $144=2 \cdot 72$), in implementing  the Osipov-Sommers-{\.Z}yczkowski formula (\ref{JointBuresHSformula}) given above with $x=\frac{1}{2}$. (For the companion Bures rebit-retrit estimation, we would have a smaller $d$, that is, 78--but given our Bures two-rebit analysis above (sec.~\ref{BuresTwoRebit}), we were not optimistic in being able to advance a possible exact value.) In Fig.~\ref{fig:BuresQubitQutrit} we show a (scaled) plot of our corresponding computations. The estimates--recorded at intervals of one million--are in general agreement with the reported value of Khvedelidze and Rogojin. The last  value (after 3,174 million iterations) was $\frac{1479997}{1058000000}= \approx 0.001398863$. This can be well-fitted by  $\frac{1}{715} =\frac{1}{5 \cdot 11 \cdot 13} \approx 0.00139860$.
\subsection{Higher-Dimensional Bures Analyses}
\begin{figure}
    \centering
    \includegraphics{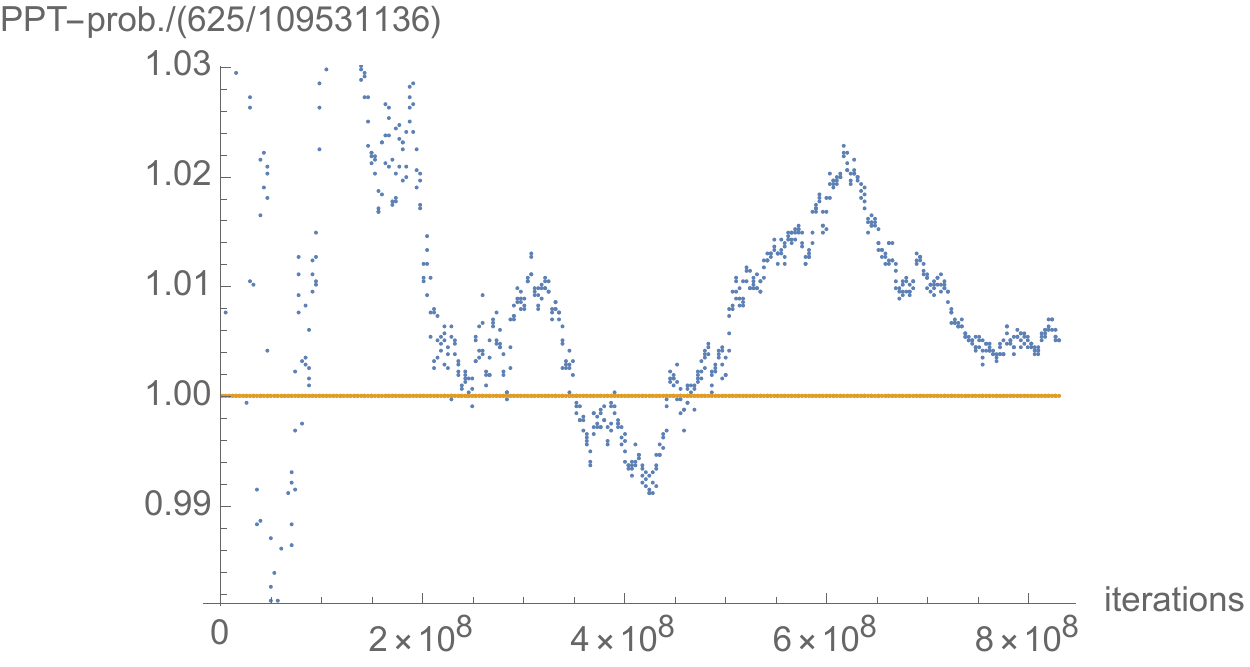}
    \caption{Qubit-qudit ($2 \times 4$) Bures PPT-probability estimates--divided by $\frac{625}{109531136}=\frac{25^2}{2^{12} \cdot 11^2 \cdot 13 \cdot 17}$--as a function of the number of iterations of the quasirandom procedure, using $\alpha_0=\frac{1}{2}$. Estimates are recorded at intervals of one million iterations.}
    \label{fig:BuresQubitQudit}
\end{figure}

\begin{figure}
    \centering
    \includegraphics{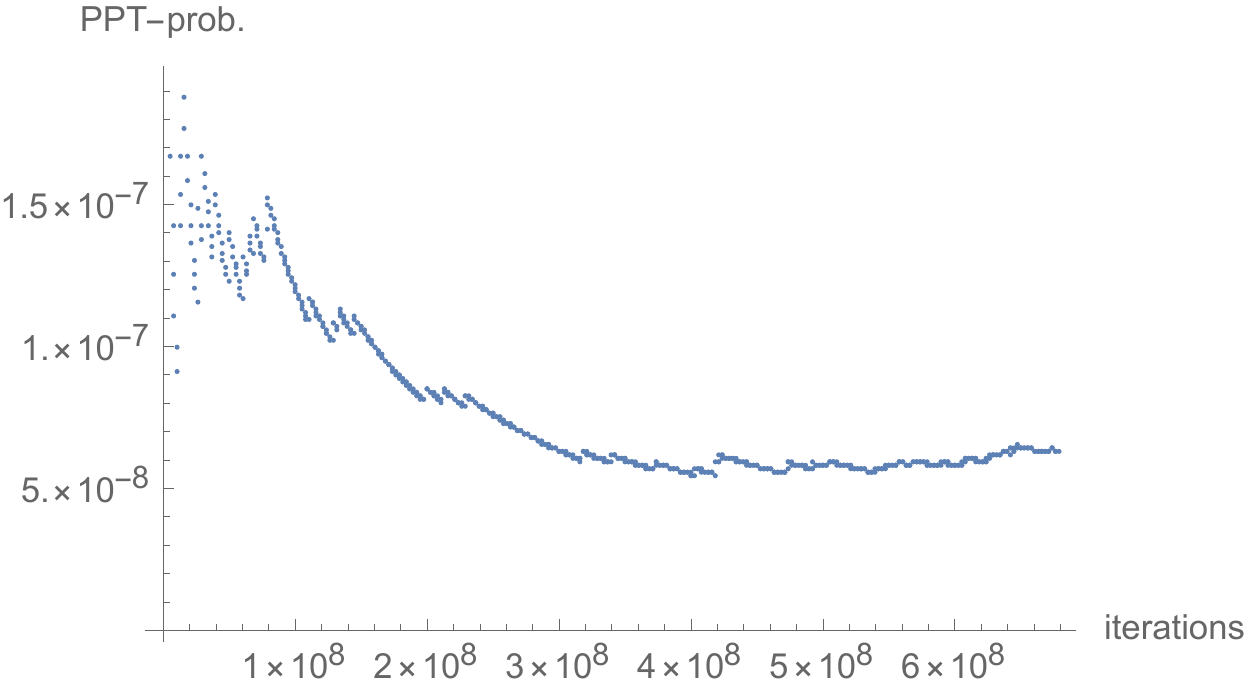}
    \caption{Two-qutrit Bures PPT-probability estimates as a function of the number of iterations of the quasirandom procedure, using $\alpha_0=\frac{1}{2}$. Estimates are recorded at intervals of one million iterations.}
    \label{fig:BuresQubitQudit2x5}
    \end{figure}

To estimate the Bures qubit-qudit ("ququart") bipartite ($2 \times 4$) PPT-probability, we employed a 256-dimensional quasirandom sequence, obtaining 4,760 PPT density matrices in 830 million realizations, yielding an estimated probability of $5.7349398 \cdot 10^{-6}$. An interesting candidate for a possible corresponding exact value is 
$\frac{625}{109531136}=\frac{25^2}{2^{12} \cdot 11^2 \cdot 13 \cdot 17} \approx 5.70614003 \cdot 10^{-6}$ (Fig.~\ref{fig:BuresQubitQudit}).
For the Bures two-qutrit scenario, employing a 324-dimensional sequence, only 43 PPT density matrices were generated in 678 million realizations, yielding an estimate of  $6.3421829 \cdot 10^{-8}$ (Fig.~\ref{fig:BuresQubitQudit2x5}). (It would be of interest to relate this last very small PPT-probability estimation to the asymptotic analyses of Aubrun and Szarek \cite{aubrun2017alice}, as well as Ye \cite{ye2009bures}.)
\section{Higher-Dimensional Hilbert-Schmidt Analyses} 
Further, in
\cite[sec. 3,5]{slater2018extensions}, we had suggested Hilbert-Schmidt PPT-probability hypotheses for the $2 \times 4$ and $2 \times 5$ qubit-qudit systems of $\frac{16}{12375} =\frac{4^2}{3^2 \cdot 5^3 \cdot 11} \approx 0.001292929$ and $\frac{125}{4790016} = \frac{5^3}{2^8 \cdot 3^5 \cdot 5  \cdot 7 \cdot 11} \approx 0.0000260959$, and $\frac{201}{8192} = \frac{3 \cdot 67}{2^{13}} \approx 0.0245361$ 
and $\frac{29058}{9765625}= \frac{2 \cdot 3 \cdot 29 \cdot 167}{5^{10}} \approx 0.00297554$ for their respective rebit-redit analogues. 

\begin{figure}
    \centering
    \includegraphics{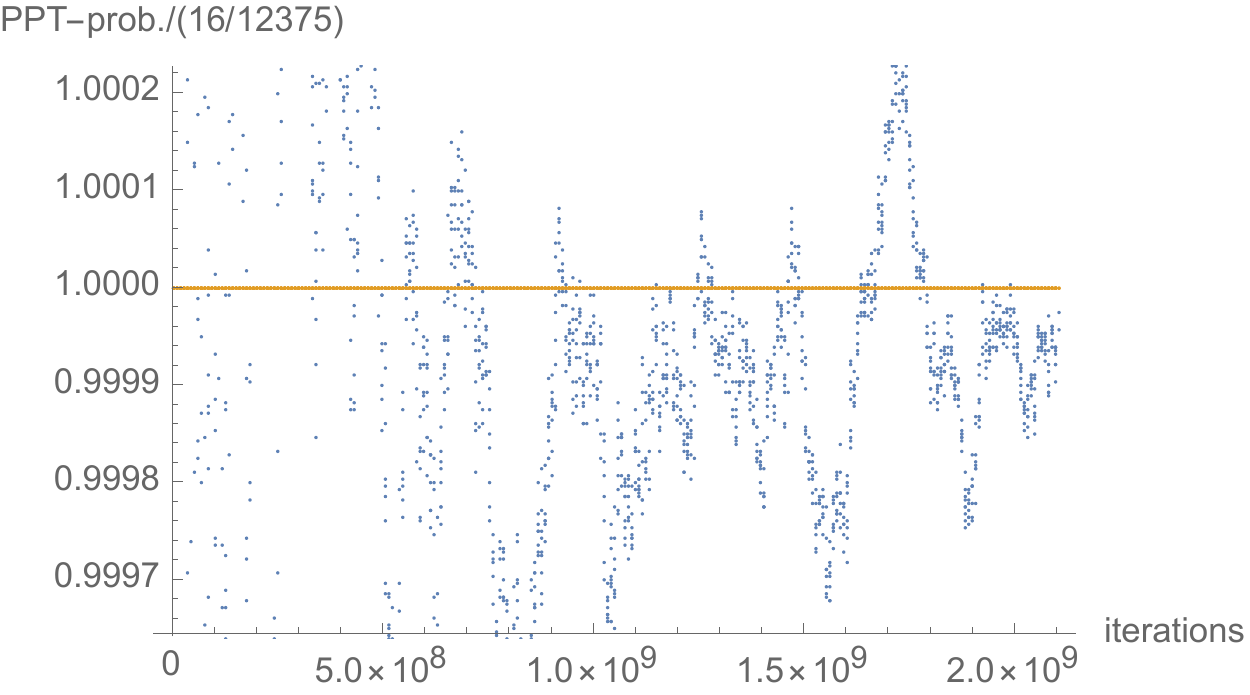}
    \caption{Qubit-qudit ($2 \times 4$) Hilbert-Schmidt PPT-probability estimates--divided by the conjectured value $\frac{16}{12375} =\frac{4^2}{3^2 \cdot 5^3 \cdot 11} \approx 0.001292929$--as a function of the number of iterations of the quasirandom procedure, using $\alpha_0=\frac{1}{2}$. Estimates are recorded at intervals of one million iterations.}
    \label{fig:HSQubitQudit}
\end{figure}

\begin{figure}
    \centering
    \includegraphics{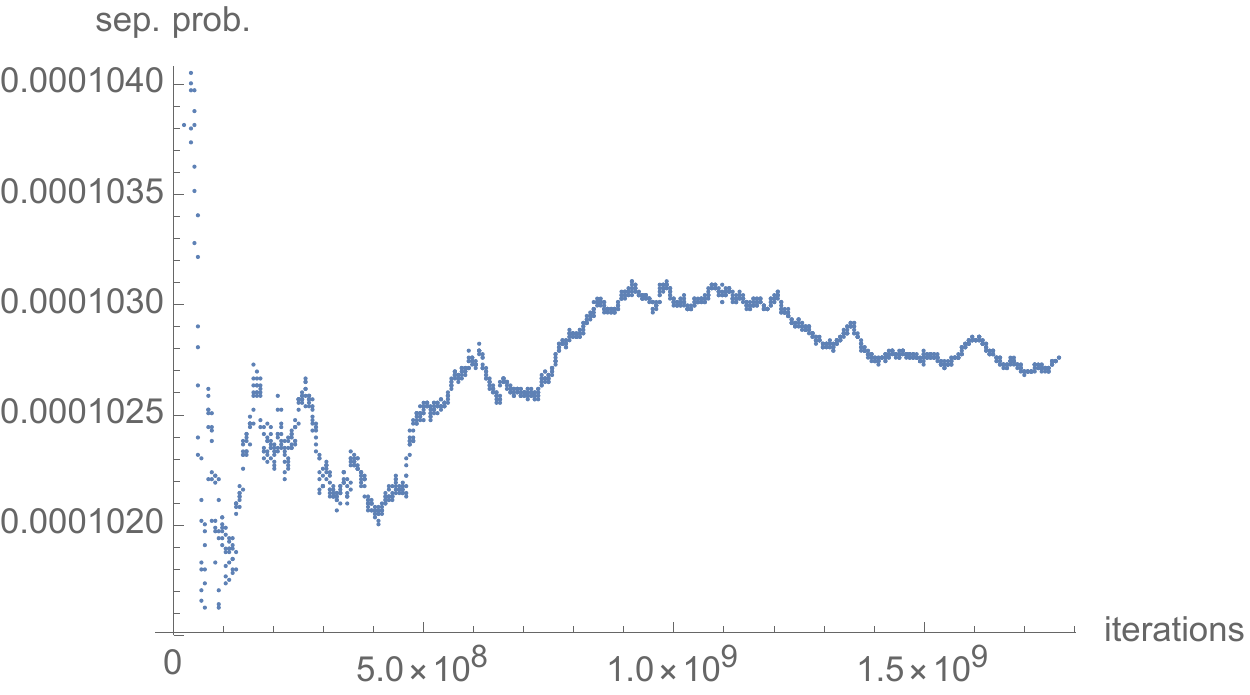}
    \caption{Two-qutrit Hilbert-Schmidt PPT-probability estimates as a function of the number of iterations of the quasirandom procedure, using $\alpha_0=\frac{1}{2}$. Estimates are recorded at intervals of one million iterations.}
    \label{fig:HSTwoQutrit}
\end{figure}

For the Hilbert-Schmidt $2 \times 4$ qubit-qudit and two-qutrit scenarios, using the quasirandom procedure introduced by Martin Roberts \cite{Roberts}, we have obtained PPT-probability estimates of 0.0012928963 and 0.00010275452 based on 2,104 and 1,768 million iterations, respectively (Figs.~\ref{fig:HSQubitQudit} and \ref{fig:HSTwoQutrit}). 
In further regard to Hilbert-Schmidt two-{\it qutrit} probabilities, an estimate of 0.00010218 based on 100 million random realizations was reported in sec. III.A of ``Invariance of Bipartite Separability and 
PPT-Probabilities over Casimir Invariants of Reduced States'' \cite{slater2016invariance}. (An intriguing possible corresponding exact value is $\frac{323}{3161088}=\frac{17 \cdot 19}{2^{10} \cdot 3^2 \cdot 7^3} \approx 0.000102180009$--or $\frac{11}{107653}=\frac{11}{7^2 \cdot 13^3} \approx 0.000102180153$.)

\subsection{Use of realignment criterion for (bound-)entanglment estimations}
Also, in an auxiliary  $2 \times 4$ qubit-qudit analysis, based on 795 million iterations, 
use of the realignment criterion \cite{chen2002matrix} yielded an 
estimate of 0.000234478 for the bound-entangled probability  and 0.942343 (conjecturally, $\frac{589}{625}=\frac{17 \cdot 31}{5^4} \approx 0.9424$) for the entanglement probability, in general. (The PPT-probability was, once again, well fitted--to almost five decimal places--by $\frac{16}{12375}$.) In that analysis, we were not able to detect any finite probability at all of genuinely tripartite entanglement using the Greenberger-Horne-Zeilinger test set out in Example 3 in \cite{bae2018entanglement}. 
However, in a parallel two-qutrit study, the realignment test for entanglement was not passed by any randomly generated states (cf. \cite[sec. IV]{gabdulin2019investigating}). 
\subsection{The question of optimality of 64D low-discrepancy sequences}
It may be of interest to the reader to here include a response of Martin Roberts to a query as to whether
to calculate a 64D integral, it is optimal or not to use a 64D low-discrepancy sequence, as employed above in the two-qubit case.
Roberts interestingly replied: ``It depends. In theory, the convergence rate of simple random sampling is O(1/n), whereas for low discrepancy sequences it is O($\frac{\log(N)^d}{N}$). The $\log(N)^d$ term implies that in theory for some large D, and very large N, the convergence rate of quasirandom sequences is inferior to simple random sampling. However, the classic Big O notation ignores two  things, which in practice are crucial. 
(i) Big O notation is for $N \rightarrow \infty$. For finite N, the constants of proportionality play a big role in determining which one is more efficient. 
(ii) it has been found that for many high dimensional integrals (especially the finance, computer vision, and natural language processing)  although they may outwardly look like high dimensional functions they are in fact really relatively low dimensional problems embedded in a high dimensional manifold. Therefore the pragmatic D in the above expression, is really the 'intrinsic' D. This is why finance options-pricing which is based on integrations over a few hundred dimensions are still more efficient with quasirandom sampling."

\section{Concluding Remarks}
We should stress that the problem of {\it formally deriving} the Bures two-rebit and two-qubit separability probabilities, and, thus, testing the candidate value ($\frac{25}{341}$) advanced here (Fig.~\ref{fig:twoqubitplot}), certainly currently seems intractable--even, it would seem, in the pioneering framework of Lovas and Andai \cite{lovas2017invariance}. (Perhaps some formal advances can be made, in such regards,  with respect to $X$-states [cf. \cite{xiong2017geometric}].)

Let us note that the ``master Lovas-Andai" formula for {\it generalized} two-qubit Hilbert-Schmidt ($k=0$) separability probabilities reported in \cite[sec. VIII.A]{slater2017master}
\begin{equation} \label{MasterFormula}
\tilde{\chi}_{d,0}(\varepsilon) \equiv  \tilde{\chi_d}(\varepsilon)= \frac{\varepsilon ^d \Gamma (d+1)^3 \,
   _3\tilde{F}_2\left(-\frac{d}{2},\frac{d}{2},d;\frac{d}{2}+1,\frac{3
   d}{2}+1;\varepsilon ^2\right)}{\Gamma \left(\frac{d}{2}+1\right)^2},  
\end{equation}
($\varepsilon$ being a singular-value ratio, and $d$--{\it not} the quasirandom dimension parameter--the random-matrix Dyson index)
has been recently extended  to apply to the still  more general class of ``induced measures" \cite{zyczkowski2001induced}, giving expressions for $\chi_{d,k}(\varepsilon)$ \cite{slater2018extensions}. (Also, we have sought to develop an alternative framework to that of Lovas and Andai, in the context of 
``Slater separability functions'', but not yet fully successfully \cite{LovasAndaiAlternative1,LovasAndaiAlternative2}.)

As specific illustrations here of (\ref{MasterFormula})--with the assistance of C. Dunkl--are the formulas \cite[sec. B.3.c]{slater2018extensions}--with $z=\varepsilon^2$--for $\chi_{2,k},\chi_{4.k}$ and
$\chi_{6,k}$:%
\begin{align*}
\chi_{2,k}\left(  z\right)   &  =1+\left(  1-z\right)  ^{k+1}\left(
-1+\frac{1}{k+3}z\right)  ,\\
\chi_{4.k}\left(  z\right)   &  =1+\left(  1-z\right)  ^{k+1}\left(
-1-\left(  k+1\right)  z+\frac{2\left(  2k^{2}+14k+21\right)  }{\left(
k+5\right)  \left(  k+6\right)  }z^{2}-\frac{6\left(  k+3\right)  }{\left(
k+6\right)  \left(  k+7\right)  }z^{3}\right)  ,\\
\chi_{6,k}\left(  z\right)   &  =1+\left(  1-z\right)  ^{k+1}\{-1-\left(
k+1\right)  z-\frac{\left(  k+1\right)  \left(  k+2\right)  }{2}z^{2}\\
&  +\frac{3\left(  3k^{4}+60k^{3}+432k^{2}+1230k+1264\right)  }{2\left(
k+7\right)  \left(  k+8\right)  \left(  k+9\right)  }z^{3}-\frac{6\left(
k+4)(3k^{2}+33k+80\right)  }{\left(  k+8\right)  \left(  k+9\right)  \left(
k+10\right)  }z^{4}\\
&  +\frac{30\left(  k+4\right)  \left(  k+5\right)  }{\left(  k+9\right)
\left(  k+10\right)  \left(  k+11\right)  }z^{5}\}.
\end{align*}

In section 4 of their recent study \cite{lovas2017invariance}, Lovas and Andai extended their analyses from one involving the (non-monotone \cite{ozawa2000entanglement}) Hilbert-Schmidt measure to
one based on the operator monotone function $\sqrt{x}$. They were able to conclude (for the case $d=1$ [a Dyson-type random-matrix index]) that the applicable ``separability function" in this case, 
$\tilde{\eta}_d(\varepsilon)$, 
is precisely the same as the Hilbert-Schmidt counterpart  $\tilde{\chi}_d(\varepsilon)$.

Now, quite strikingly, we obtained \cite{slater2017master}, using this function, for the two-qubit ($d=2$) analysis, the ratio of
$\frac{\pi ^2}{2}-\frac{128}{27}$ to $\frac{\pi^2}{2}$, that is,
\begin{equation}
\mathcal{P}_{sep.\sqrt{x}}(\mathbb{C})  = 1-\frac{256}{27 \pi ^2} =1 -\frac{4^4}{3^3 \pi^2} \approx 0.0393251.
\end{equation}
(We observe that such results--as with the Hilbert-Schmidt value of 
$\frac{8}{33}$--interestingly appear to reach their most simple/elegant in the [standard, 15-dimensional] two-{\it qubit} setting, where the off-diagonal entries of the density matrix are, in general, complex-valued.)

Lovas and Andai have shown that the two-rebit separability probability based on the operator monotone function $\sqrt{x}$ is approximately 0.26223001318, asserting that the integrand can be evaluate[d] only numerically". Nevertheless, we  
investigated--so far, rather not too productively, as with the Bures two-rebit estimate 0.157096234 above (sec.~\ref{BuresTwoRebit})--the possibility of finding an exact, underlying value for this statistic.  (Our investigation, in this regard, is reported in \cite{StackExchange}. It involved first performing a series expansion of the elliptic and hypergeometric functions in their integrand. We were able to then integrate this series expansion, but only over a restricted range--rather than $[0,\infty]$--of the two indices. Numerical summation over this restricted set yielded a value of only 0.0042727 [reported in \cite{StackExchange}] {\it vs.} 0.26223001318.)

It would be of substantial interest to compare/contrast the relative merits of our  quasirandom estimations above of the two-rebit and two-qubit Bures separability probabilities in the 36- and 64-dimensional settings employed with earlier studies
(largely involving Euler-angle parameterizations of $4 \times 4$ density matrices \cite{tilma2002parametrization}), in which 9- and 15-dimensional integration problems were addressed \cite{slater2005silver,slater2009eigenvalues} (cf. \cite{maziero2015random}). In the higher-dimensional frameworks used here, the integrands are effectively unity, with each randomly generated matrix being effectively assigned equal weight, while not so in the other cases indicated. In \cite{ExperimentalData}, we asked the question ``Can `experimental data from a quantum computer' be used to test separability probability conjectures?'', following the analyses of Smart, Schuster and Mazziotti in their article \cite{ssm}, ``Experimental data from a quantum computer verifies the generalized Pauli exclusion principle'', in which 
``quantum many-fermion states are randomly prepared on the quantum computer and tested for constraint violations''.

So, in brief summary, let us state that at this stage of our continuing investigations, it appears that we have a set of three exact-valued measure-dependent two-qubit separability probabilities ($\frac{8}{33}$ [Hilbert-Schmidt], $1-\frac{256}{27 \pi^2}$ [operator monotone $\sqrt{x}$], $\frac{25}{341}$ [Bures--minimal monotone $\frac{1+x}{2}$]), but only one two-rebit one ($\frac{29}{64}$ [Hilbert-Schmidt]). 

The [apparent lesser than $\frac{25}{341}$] separability probabilities for other members--Kubo-Mori, Wigner-Yanase,\ldots--of the monotone family have been estimated in \cite{slater2005silver}--cf. \cite{singh2014relative,batle2014geometric}. But since there is, at present, no apparent mechanism available for generating density matrices random with respect to such measures [cf. \cite[sec. V.B]{puchala2011probability} in regard to superfidelity], the quasirandom procedure seems unavailable for them. (Also the use of measures that are non-monotone in nature--in addition to the well-studied  Hilbert-Schmidt one--would be of interest, for example, the Monge \cite{zyczkowski2001monge} and Husimi \cite{slater2006quantum,rexiti2018volume} measures.) However, separability/PPT-probabilities can be so analyzed for the class of induced measures \cite{zyczkowski2001induced}. 

Let us pose the following problem: construct a function $f$ that yields the separability probabilities associated with the monotone metrics. That is, we would have (the Bures case) $f(\frac{1+t}{2})=\frac{25}{341} =0.0733138$, $f(\sqrt{t}) =1 -\frac{256}{27 \pi^2}=0.0393251$ and  $f(\frac{2 t}{1+t}) =0$. Additionally, $f(\frac{t^{(t-1)}}{e})  \approx 0.0609965$, $f(\frac{1}{4} \left(\sqrt{t}+1\right)^2) \approx 0.0503391$ and $f(\frac{(t-1)}{\log{t}}) \approx .0346801$ \cite[Tab. II]{slater2005silver} and also $f(\frac{1+6 t +t^2}{4 +4 t}) \approx 0.0475438$ \cite[Tab. I]{slater2005silver}.

\begin{acknowledgements}
This research was supported by the National Science Foundation under Grant No. NSF PHY-1748958.
\end{acknowledgements}

\bibliography{main}

\end{document}